\documentclass[12pt,preprint]{aastex}

\def\sqig{$\sim$}
\def\sun{$_\odot$}
\def\degrees{$^{\circ}$}

\def\ditto{\texttt{"}}
\def\source{EXO\,1722-363}
\def\src{EXO\,1722-363}

\begin{document}
\title{
Rossi X-ray Timing Explorer Observations of the X-ray Pulsar EXO 1722-363
- a Candidate Eclipsing Supergiant System}

\author{Robin H.D. Corbet\altaffilmark{1},
Craig B. Markwardt\altaffilmark{2},
and Jean H. Swank}

\affil{X-ray Astrophysics Laboratory, Code 662,\\
NASA/Goddard Space Flight Center, Greenbelt, MD 20771}
\altaffiltext{1}{Universities Space Research Association}
\email{corbet@gsfc.nasa.gov}
\altaffiltext{2}{Department of Astronomy, University of Maryland}

\begin{abstract}

Observations made of the X-ray pulsar \src\ using the Proportional Counter
Array and All Sky Monitor on
board the Rossi X-ray Timing Explorer reveal the orbital period of
this system to be 9.741 $\pm$ 0.004 d from periodic changes in the source flux. The
detection of eclipses, together with the values of
the pulse and orbital periods, suggest that this source consists of a
neutron star accreting from the stellar wind of an early spectral type
supergiant companion. Pulse timing measurements were also obtained but
do not strongly constrain the system parameters. The X-ray spectra
can be well fitted with a model consisting of a power law with a high
energy cutoff and, for some spectra, a blackbody 
component with a temperature of approximately 0.85 keV.

\end{abstract}

\section{Introduction}

The X-ray source \src\ (X1722-36) was discovered during Galactic plane
observations made with EXOSAT (Warwick et al. 1988).  Observations
made with Ginga by Tawara et al. (1989) showed the presence of 413.9s
pulsations and also revealed variability
on a timescale of hours.
Tawara et al. (1989) proposed that \src\ was a system consisting of a neutron
star accreting material from a Be star.
Additional pulse timing observations were made with Ginga
by Takeuchi, Koyama \& Warwick (1990) who derived lower limits
on the orbital period and mass of the primary star of 9 d and 15 M\sun\
respectively. Takeuchi et al. (1990) also observed X-ray flaring activity
on time scales down to a few hours.

In this paper we present the results of extended monitoring
scan observations
made with the Proportional Counter Array (PCA) on
board the Rossi X-ray Timing Explorer (RXTE) satellite which reveal
the orbital period of the system. A set of pointed
observations was also obtained which provide spectral and
pulse timing information. In addition, utilizing a refined source
position obtained by the INTEGRAL satellite (Lutovinov, Revnivtsev
\& Molkov 2003,
Revnivtsev et al. 2004)
we have obtained a long term light curve with the RXTE All-Sky Monitor (ASM).
The source properties suggest that this system consists
of a neutron star accreting from the stellar wind of an early spectral
type supergiant companion.

\section{Observations}

In this paper we present the results of observations of \src\ that have
been made with two of the instruments on board RXTE (Bradt, Rothschild, \&
Swank 1983): the PCA and the ASM.

\subsection{RXTE PCA}

The PCA is described in detail by Jahoda et al. (1996).  This instrument
consists of five, nearly identical, Proportional Counter
Units (PCUs) sensitive to X-rays with energies between 2 - 60 keV with
a total effective area of \sqig6500 cm$^2$. The PCUs each have a
multi-anode xenon-filled volume, with a front propane volume which is
primarily used for background rejection.  
The Crab produces 13,000
counts/s for the entire PCA across the complete energy band.  The PCA spectral resolution at 6 keV is approximately 18\%
and the field of view is 1\degrees\ full width half maximum (FWHM).
Not all of the PCUs are always operated during an
observation and corrections have been made in our
analysis, where necessary, for the varying number of
PCUs operating at any one time.
Data extraction, including background subtraction, followed
standard procedures for RXTE PCA analysis described by the RXTE Guest Observer Facility
\footnote{http://heasarc.gsfc.nasa.gov/docs/xte/}. 

PCA observations of \src\ were obtained in two different ways:\\
(i) A series of scans across the Galactic center region has been performed
since 1999 (Swank \& Markwardt 2001) and data from the interval February 1999 to October 2003 
are presented here.  The observations consist
of raster scans of a rectangular region approximately 16\degrees $\times$ 16\degrees\
surrounding the Galactic center made on a twice weekly basis excluding November,
December, January and June when Sun constraints prevent observations.
These scans are optimized to detect faint sources and so light curves
are constructed using only the top layer of the PCA and cover the energy range
2 to 10 keV. With these parameters the Crab produces 10,400 counts/s for the sum
of all 5 PCUs. 
Individual source
count rates are modulated by the PCA collimator as they pass into and
out of the field of view.  Light curves, corrected for
non-source background, are fitted to a
model of known sources plus a model of
unresolved emission from the Galactic ridge, convolved with the collimator response
function.  The nominal $1\sigma$ sensitivity to variations is
approximately 0.5--1 mCrab in the regions more than a few degrees
from the galactic center (\src\ has Galactic coordinates of $l$ = 351.5\degrees,  $b$ = -0.4\degrees).
The light curve of \src\ derived from the PCA scans is shown in
Fig. 1
\\
(ii) Pointed observations of
\src\ were obtained between 1998 October 23 to November 9 pointed at
the {\em EXOSAT} derived coordinates
of R.A. = $17^h 25^m 55.44^s$, decl. = $-36$\degrees\ 24\arcmin\ 39.6\arcsec, for  proposal number 30142 (PI: P. Saraswat).
Additional pointed observations
were obtained, on 2003 August 22, at the 
coordinates obtained by Lutovinov et al. (2003)
of R.A. = $17^h 25^m 24.00^s$, decl. = $-36$\degrees\ 18\arcmin\ 00\arcsec,
which are offset 0.15\degrees\ from the {\em EXOSAT} coordinates,
and on 2003 August 29, again at the {\em EXOSAT} position. These were the
public target of opportunity observations from
proposal number 80424,
carried out in response to
the report of {\em INTEGRAL} observations of activity. Another
observation, of XTE\,J1723-376 for proposal number 40705
(PI: W. Cui), occurred in 1999,
during an eclipse of \src. XTE\,J1723-376 was in quiescence,
as indicated by ASM and the PCA bulge scans, during the observations
of \src.
Observations
were interrupted because
of instrumental constraints such as Earth occultations of
the source and passages through the South Atlantic Anomaly when
the instruments are not operated, and because observations of
other sources were also undertaken during these periods. The
resulting total effective exposure time was 50.9 ks. 
All 5 PCUs were operating during the 1998 observations and 3 PCUs were
operating during the 2003 observations.
The light curve from the pointed observations used in the analysis presented
here includes photons in the energy range of approximately 2.0 to 30 keV
taken from all three layers of the PCA.
The light curve from the pointed PCA observations of \src\
obtained in 1998 is shown in Fig. 2.

\subsection{RXTE ASM}

The ASM (Levine et al. 1996) consists of three similar
Scanning Shadow Cameras, sensitive to X-rays in an energy band of
approximately 2-12 keV, which perform sets of 90 second pointed
observations (``dwells'') so as to cover \sqig80\% of the sky every
\sqig90 minutes.  
The Crab produces approximately 75
counts/s in the ASM over the entire energy range. Observations
of blank field regions away from the Galactic center suggest that
background subtraction may produce a systematic uncertainty of about 0.1
counts/s (Remillard \& Levine 1997). To extract a light curve
from ASM observations it is necessary to have an accurate source
location and the coordinates
reported by Lutovinov et al. (2003), which have a quoted error
of 2\arcmin, were used to create the light curve
of \src. We note Lutovinov et al. (2004) give
coordinates that are 3\arcmin\ from this location; this
difference is too small to make a large difference to the ASM light
curve.
The ASM light curve of \src\
considered here covers approximately 9 years. 
The ASM data were filtered by excluding all dwells
where the modeled background in the lowest energy band was
greater than 10 counts/s. This procedure helps to exclude points
where the data are contaminated by solar X-rays.

\section{Results}

\subsection{RXTE PCA}

\subsubsection{PCA Scans - The Light Curve and the Orbital Period}

Using the long term light curve obtained from the PCA scan observations (Fig. 1)
a power spectrum was calculated. The results of this are shown in Fig. 3. 
A highly significant peak is seen at a frequency corresponding
to a period of approximately 9.7 days. The orbital period determination
was refined by undertaking a sine wave fit to the light curve.
This procedure yielded a period of 9.741 $\pm$ 0.004 days
with epoch of minimum flux of MJD 51218.97 $\pm$ 0.22.
The PCA scan light curve folded on this period is shown in Fig. 4.
We note the presence of a deep minimum, consistent with an eclipse,
at phase 0.

In order to quantify this eclipse we fitted the folded light curve
with a simple eclipse model consisting of constant flux outside eclipse,
constant flux during full eclipse, and linear transitions between these levels
during eclipse ingress and egress. The results of this fit are given
in Table 1. As the eclipse profile fit gives a more precise value for the center
of the minimum than does the sine wave fit we use the eclipse fit to define
``T0'' (eclipse center) in our following analysis.

From the eclipse profile fit we find a nominal count rate of -0.3 $\pm$ 0.1 counts/s/PCU during the eclipse.
That is, the count rate is consistent with a total eclipse within the errors with a slight
systematic underestimate from the fitting process.


We can extrapolate our ephemeris back to the two sets of Ginga
observations. During the observations reported by Tawara et al. (1989)
the flux of \src\ was observed to decline to a very low level on
1987 October 10 (MJD 47078). Our ephemeris gives a phase
of $\phi$ = 0.93 $\pm$ 0.17 for the end of this observation.
In the light curve presented by Takeuchi et al. (1990) a very low
flux was observed on 1988 April 3 (MJD 47254). Our ephemeris
gives $\phi$ = 0.92 $\pm$ 0.17. Thus both Ginga observations of low
flux from \src\ are consistent with these events being associated
with an eclipse.

\subsubsection{PCA Pointed Observations - the Light Curve and Orbital Modulation of
Pulse Arrival Times}

The light curve from the pointed PCA observations of \src\ obtained in 1998 is
shown in Fig. 2. There is considerable source variability and pulsations
are clearly detected at most times.
During the interval
MJD 51122.0 to 51122.16
the source flux is observed to be constant at a low level of about
4.9 counts/s/PCU.
This corresponds to phases of 0.006 to 0.022 using our ephemeris 
from the scan observations and, during this interval, no pulsations are detectable.
The source is thus likely to be completely in eclipse at this time. 
\src\ is located close to the Galactic center at $l$ = 351.5\degrees,
$b$ = -0.3\degrees and significant emission from the Galactic ridge (e.g.
Valinia \& Marshall 1998),
and perhaps other sources, is to be expected within the PCA field of view.
The non-zero
count rate found in the pointed
observations is consistent with the modeled galactic ridge flux used
in determining source flux seen in the scans. 
Between MJD 51113.1 to 51113.3 (phases 0.092 to 0.113) the source is seen to be
increasing from a low flux level of
about 4.9 counts/s/PCU (Fig. 5), consistent with the observation of an eclipse egress. 

Pulse arrival times were obtained from this light curve by
means of a cross-correlation procedure. 
The sparse orbital coverage severely limits
the constraints which the timing data  can  place on the orbital parameters.
We fitted a circular orbit with no spin period change, 
and with the orbital period  fixed at the value obtained
from our PCA scan data.
We conducted a grid search to find the orbital solution with
the lowest $\chi^2$ for the pulse arrival times.
The solution is given in Table 2 and plotted in Fig. 6.
The best fit for T90 (the same as the eclipse center for a circular
orbit) agrees with the center of the eclipse in Table 1. 
We note that the value for $a\ $sin\ $i$ is
consistent with the constraints obtained by Takeuchi et al. (1990) for
an orbital period of 9.741 d.
The overall pulse profile, corrected for the  orbital
solution is plotted in Fig. 7.

\subsubsection{PCA Pointed Observations - Spectroscopy}

For spectral analysis we divided the pointed PCA observations
into 18 separate accumulated spectra as specified in Table 3.
One spectrum was accumulated for
each ``Good Time Interval'', i.e. each continuous data collection period
between source occultations, South Atlantic Anomaly passages or
observations of other sources. Response matrices were created using
version 10.1 of the program ``pcarmf''.

The light curve from the pointed PCA observations (Section 3.1.2, Fig. 2),
in particular the significant flux detected during eclipse in contrast to
the scan observation
indicates significant Galactic ridge contamination. We therefore first
fitted the eclipse spectra (numbers 11, 12, and 13) to determine the
level of this contribution. Following Valinia \& Marshall (1998)
we fitted a model consisting of a Raymond-Smith plasma and power-law components
and 
this model was found to give a good fit. The spectral fit
is shown in Fig. 8 and the best fit parameters are given
in Table 4. This spectrum was included as a fixed
component in the models for the spectra
of other phases.

The spectra of X-ray pulsars are commonly fit with a model
consisting of an absorbed power
law spectrum with a high energy cutoff and, in some cases, an iron line near
6.4 keV (e.g. White, Swank \& Holt, 1983).
Such a model
was employed by Takeuchi et al. (1990) to fit the spectrum of
\src, while Tawara et al. (1989) used a simpler model with no
high-energy cutoff.
For the PCA data (fitting the 2-40 keV range) we find, however,
that not all the RXTE spectra of \src\ can be satisfactorily fit
with the usual cutoff power-law spectrum. With experimentation,
we found that adding a low energy (\sqig 0.85 keV) blackbody
component gave satisfactory fits in those cases, with the exception of spectrum number
1, which required additional components of an iron line and an additional
unabsorbed blackbody.
Only for spectrum number 1  was an
iron line clearly required and so
we did not include this component in the final fits of the other spectra.
The resulting spectral parameters
are listed
in Tables 5 and 6 and the spectra and fits are
shown in Fig. 9. 
We note that the hydrogen column density obtained appears to
be relatively constant with the exception of the three spectra
obtained during eclipse egress (5, 6, \& 7) and the one (17) closest to ingress; for these increased
absorption is found. 
The intrinsic spectrum does vary, but not dramatically, and the source is too faint (0.5-5 mCrab) 
and the observations too incomplete to draw conclusions about the variations. Although spectrum number 1 is the only spectrum
that requires additional model components, we find no evidence
for any problem, such as excess background,
with this particular spectrum. 
Since the PCA has a 1\degrees\ FWHM field of view, it is possible
that a contaminating source present in the field had a flare
that affected only this spectrum
and not spectrum number 2 which was obtained shortly after
spectrum number 1.
Spectrum 17 shows some small but possibly systematic residuals in
the vicinity of 6.4 keV. This may be an indication that the background
was somewhat different for this slightly
different pointing position.



\subsection{RXTE ASM - Light Curve and Orbital Modulation}

The mean count rate from \src\ measured with the ASM was 0.3 counts/s.  We calculated
the power spectrum of the RXTE ASM light curve of \src\ and found that, although a small peak
was present in the power spectrum
near the orbital period
at 9.738 $\pm$ 0.02d (HWHM), it would not have been possible a
priori to determine the orbital period of this system from the ASM data
alone. This peak is not of statistical significance
and is far from the strongest peak in the power spectrum.
In Fig. 10 we show the ASM data folded on the orbital period found from the
PCA scan data. It
can be seen that the ASM data show a very similar overall modulation to
that observed with the PCA and that the phasing of flux minimum is
consistent between the two instruments. We can thus have reasonable confidence
that the ASM is indeed measuring flux variability from \src\ in spite of the lack
of a significant peak in the power spectrum.
The quality of light curves that can be obtained with the ASM
depends on the accuracy with which a source position is known.
We note that the INTEGRAL coordinates given by Revnivtsev et
al. (2004), 
R.A. = $17^h 25^m 09.00^s$, decl. = $-36$\degrees 16\arcmin 48\arcsec\
differ by approximately 3 arc minutes from those given by Lutovinov et al. (2003)
which were used in the production of the ASM light curve for \src.

\section{Discussion}

X-ray pulsars are generally high mass systems in which the primary
is an O or B type star. This is the case
for \src\ as shown by the large mass function obtained from
the pulse timing measurements of Takeuchi et al. (1990). 
High mass X-ray binaries can be broadly divided into
those systems where the primary star is a supergiant and those
where the primary is a luminosity class III - V Be star.
Several pieces of evidence indicate that \src\ has a supergiant
primary. The detection of an eclipse indicates
a substantial size for the primary star. The probability of
a Be star system showing an eclipse is very low due to the small size of
the mass donor and, even if an eclipse were to be observed in
such a system it would only have a very short duration.
In addition, the values
of the pulse and orbital periods are also typical for a supergiant
system (Corbet 1986). 
Our orbital solution (Fig. 6) has an amplitude of
111.1 $\pm$ 0.7 s, which corresponds to a mass function of
15.4 $\pm$ 0.1 M\sun. This mass function is comparable with that
found for several other supergiant
systems (e.g. van Paradijs \& McClintock 1995), but conclusive determination of system parameters awaits
pulse arrival time measurements with more complete 
orbital phase coverage.

Supergiant systems can be further divided into those systems where
the compact object accretes from the stellar wind of the primary
and those rare systems where mass accretion occurs because the primary
fills its Roche lobe. \src\ is likely to be a wind-accretion driven
system because its relatively low luminosity and long spin period
indicate only modest mass and angular momentum accretion unlike
Roche-lobe overflow powered systems (e.g. Bildsten et al. 1997).

In several cases our spectral fits require a black body component
in addition to the usual cutoff power-law.
Soft black body spectral components have been observed in several luminous X-ray
pulsars such as SMC X-1 and LMC X-4 (e.g. Paul et al. 2002, Hickox, Narayan,
\& Kallman, 2004.
and references therein). However, in other sources the black body component
typically has a temperature of \sqig 0.1 keV rather than the \sqig 0.85 keV
temperatures in our fits.

\section{Conclusion}

\src\ is found to be an eclipsing X-ray binary where the mass-donating primary
is likely to be a supergiant star. 
It would be valuable to obtain frequent pulse timing measurements of \src\
over at least one entire orbital period so that the system parameters including
mass function and eccentricity
can be precisely determined. As \src\ is an eclipsing binary
there is the promise of determining an accurate neutron star
mass for this system. To measure this it would be very valuable if an optical
or IR counterpart
can be found and the orbital radial velocity of the
counterpart measured. This identification of a counterpart will likely require
a significantly improved X-ray position over the one currently available.
Information on the mass-donor star can also be obtained from the precise
determination of eclipse ingress and egress times. The determination 
and monitoring of eclipse times would also facilitate a search for changes
in orbital period. 
An X-ray spectrum of \src\ obtained using an imaging
instrument would also enable an interpretation of the spectrum without
the problem of contamination from the Galactic ridge
and other nearby X-ray sources.

\acknowledgments
We thank R.A. Remillard
for producing the ASM light curve of \src.



\begin{table}
\caption{Photometric Orbital Parameters for \src}
\begin{center}
\begin{tabular}{lcr}
Parameter & Value & Units \\
\tableline
Period (Sine Fit) & 9.7407 $\pm$ 0.004 & days \\
Flux Minimum (Sine Fit) & 51218.97 $\pm$ 0.22 & MJD \\
Eclipse Totality Half Width & 31.8 $\pm$ 1.8 & degrees \\
Eclipse Totality Full Width & 1.7 $\pm$ 0.1 & days \\
Eclipse Totality Start & 51218.49 $\pm$ 0.08 & MJD \\
Eclipse Totality Start & 0.916 $\pm$ 0.008  & Orbital Phase \\
Eclipse Totality End & 0.088 $\pm$ 0.008 & Orbital Phase \\
Eclipse Totality End & 51220.21 $\pm$ 0.07 & MJD \\
Eclipse Center & 51219.35 $\pm$ 0.05 & MJD \\
Ingress Duration &  1.6 $\pm$ 0.1 & days \\
Egress Duration & 1.4 $\pm$ 0.1 & days \\
\tableline
\end{tabular}
\end{center}
The eclipse full and half widths are measured from the interval between
the start and end of eclipse totality.
\end{table}

\begin{table}
\caption{Illustrative Pulse Arrival Time Parameters for \src}
\begin{center}
\begin{tabular}{lcr}
Parameter & Value & Units \\
\tableline
$a\ $sin\ $i$ & 111.1 $\pm$ 0.7 & lt seconds \\
P$_{pulse}$ & 413.85528 $\pm$ 0.00001 & seconds \\
f(M) & 15.4 $\pm$ 0.1 & M\sun \\
$\chi^2_\nu$ & 1.8 & \\
T90 & 51219.350  $\pm$ 0.006 & MJD \\
\tableline
\end{tabular}
\end{center}
Note - 112 pulse arrival time fits were made with orbital period
and phase fixed to values found from photometric analysis.
\end{table}

\begin{table}
\caption{Accumulated Spectra of \src}
\begin{center}
\begin{tabular}{lrrrr}
Spectrum & Start Time &  Stop Time &  Exposure & Phase \\
Number   &            &            &    (s)    &       \\
\tableline
   1 &      9.077 &      9.107 &         2592 &      0.681 \\
  2 &      9.110 &      9.114 &          336 &      0.683 \\
  3 &      9.140 &      9.180 &         3456 &      0.688 \\
  4 &      9.207 &      9.247 &         3456 &      0.694 \\
  5 &     13.139 &     13.179 &         3456 &      0.098 \\
  6 &     13.206 &     13.246 &         3456 &      0.105 \\
  7 &     13.277 &     13.312 &         3088 &      0.112 \\
  8 &     17.937 &     17.978 &         3488 &      0.591 \\
  9 &     18.004 &     18.044 &         3536 &      0.598 \\
 10 &     18.070 &     18.103 &         2864 &      0.604 \\
 11 &     22.002 &     22.044 &         3616 &      0.008 \\
 12 &     22.068 &     22.111 &         3664 &      0.015 \\
 13 &     22.137 &     22.166 &         2480 &      0.021 \\
 14 &     26.068 &     26.111 &         3760 &      0.426 \\
 15 &     26.137 &     26.178 &         3520 &      0.433 \\
 16 &     26.208 &     26.245 &         3184 &      0.440 \\
 17 &   1773.206 &   1773.217 &          944 &      0.789 \\
 18 &   1780.581 &   1780.603 &         1920 &      0.546 \\

\tableline
\end{tabular}
\end{center}
Start and stop times are in units of MJD - 51100. Phase refers to the
center of the exposure. The pointing position for position 17 differs
from the other spectra. Spectra 11, 12, and 13 were obtained during an eclipse
and spectrum 5 appears to be from close to the start of eclipse egress.
\end{table}

\begin{table}
\caption{Eclipse Spectral Fit to \src}
\begin{center}
\begin{tabular}{lcr}
Model Component & Parameter & Value \\
\tableline
Absorption & N$_H$($\times$10$^{22}$cm$^{-2}$) & 6.6 $\pm$ 0.5 \\
Raymond-Smith & kT (keV) & 2.4 $\pm$ 0.1 \\
              & Normalization & (8.0 $\pm$ 0.7) $\times$ 10$^{-2}$ \\
Power law     & Photon Index  & 1.00 $\pm$ 0.08 \\
              & Normalization & (1.8 $\pm$ 0.6) $\times$ 10$^{-3}$ \\
\tableline
\end{tabular}
\end{center}
A simultaneous fit was made to spectra number 11, 12, and 13 over
the range 2.5 to 50 keV.
$\chi^2/\nu$ = 161/253 = 0.64. Abundances were fixed at the solar
values.
\end{table}

\begin{deluxetable}{lrrrrrrrr}
\rotate
\tablewidth{0pt}
\setlength{\tabcolsep}{0.02in}
\tablecaption{Results of Spectral Fits to \src}

\tablehead{
\colhead{\#} & \colhead{N$_H$} & \colhead{PL Index} &
\colhead{PL Norm} &  
\colhead{BB kT} & \colhead{BB Norm} & \colhead{Cutoff Energy} & \colhead{Fold Energy} & \colhead{$\chi^2_\nu$} \\
\colhead{} & \colhead{x10$^{22}$cm$^{-2}$} & \colhead{} & \colhead{$\times$10$^{-3}$} & \colhead{keV} & \colhead{$\times$10$^{-3}$} & \colhead{keV} & \colhead{keV}
}
\startdata
  1$^{\dagger}$ & 13 $\pm$ 3   & 0.50 $\pm$ 0.06 & 4 $\pm$ 1 & 0.64 $\pm$ 0.04 & 2.4 $\pm$ 1.4 & 23.0 $\pm$ 0.6 & 3.8 $\pm$ 0.95 & 0.95/ 63 \\

  2 & 13.9 $\pm$ 1.5 & 0.45 $\pm$ 0.06 & 5 $\pm $ 1 & 0.87 $\pm$ 0.06 & 1.6 $\pm$ 0.3 & 15.6 $\pm$ 0.4 & 15 $\pm$ 1 & 1.3/206 \\
  3 & 12   $\pm$ 1   &      \ditto      & 5 $\pm$  1 &      \ditto  & 1.3 $\pm$ 0.3 &     \ditto &    \ditto   &   \\
  4 & 12   $\pm$ 1   &       \ditto    & 6 $\pm$  1 &      \ditto     & 1.3 $\pm$ 0.3 &    \ditto &    \ditto  &   \\
  
  5 & 239 $\pm$ 40 & 0.7 $\pm$ 0.2 & 1.2 $\pm$ 0.7 & - & - & 21 $\pm$ 2 & 18 $\pm$ 6 & 0.55/208  \\
  6 & 162 $\pm$ 15 &    \ditto     & 3.6 $\pm$ 3.2 & -  & - &    \ditto  &    \ditto   &  \\
  7 & 95  $\pm$ 11 &    \ditto     & 4.6 $\pm$ 3.7 & -  & -  &    \ditto   &    \ditto  &  \\

  8 & 18 $\pm$ 3 & 0.33 $\pm$ 0.07 & 2.6 $\pm$ 0.7 & 0.82 $\pm$ 0.07 &0.8 $\pm$ 0.3 & 18.0 $\pm$ 0.5 & 12 $\pm$ 1 & 1.0/206 \\
  9 & 18 $\pm$ 2 & \ditto &2.5 $\pm$ 0.7 & \ditto &  0.9 $\pm$ 0.3 &  \ditto  &   \ditto  &    \\
 10 & 17 $\pm$ 2 &  \ditto & 2.3 $\pm$ 0.6& \ditto &  0.9 $\pm$ 0.4&  \ditto & \ditto &     \\
 
14 & 9 $\pm$ 1 &  0.99 $\pm$ 0.05 & 3.5 $\pm$ 0.4&  - & - &22 $\pm$ 2 & 2.4 $\pm$ 2.7 &   1.05/208\\
 15 & 6.5 $\pm$ 0.9 &     \ditto  & 2.5 $\pm$ 0.5 & -& - &  \ditto &  \ditto &       \\
 16 & 12 $\pm$ 1 & \ditto & 3.5 $\pm$ 0.6 & -& -& \ditto &  \ditto &  \\

17 & 42 $\pm$ 4& 1.4 $\pm$ 0.3 & 7.6 $\pm$0.09 & - & - & 22 $\pm$ 1 & 5 $\pm$2& 1.1/127\\
18 & 11.6 $\pm$ 0.7&  \ditto & 16 $\pm$ 1 & - & - & \ditto & \ditto &  \\ 
 
\enddata

\tablecomments{All parameter errors are 1$\sigma$ single-parameter
 confidence levels. Multiple spectra were fit simultaneously when
 they occurred close together in time and had similar behavior. In
 those cases (2-4, 5-7, 8-10, 14-16, 17-18) the $\chi^2$ is the sum for the combined fits. $^{\dagger}$ Spectrum
 1
required an unabsorbed additional black body (not shown in the table) with kT = 0.28 $\pm$ 0.01,
normalization = 8 $\pm$ 2 $\times 10^{-3}$  to fit a soft excess and an Fe
 emission line of 1.2 $\pm$ 0.6 $\times 10^{-4}$ photons s$^{-1}$ cm$^{-2}$ at
 6.4 keV. These were only significant in that spectrum.   
Orbital phase and time of each spectrum is given in
Table 3. Model normalizations follow definitions used in XSPEC (Arnaud 1996).}
\end{deluxetable}

\begin{table}
\caption{Model Fluxes from Spectral Fits to \src}
\begin{center}
\begin{tabular}{lrr}
Spectrum & Flux (2 - 10 keV) & Flux (10 - 60 keV) \\
Number     & 10$^{-10}$ ergs cm$^{-2}$ s$^{-1}$ &  10$^{-10}$ ergs cm$^{-2}$ s$^{-1}$ \\
\tableline

1 &
   1.4 &
   4.8  \\
2 &
   1.4  &
   8.7   \\
3 &
  1.4  &
  8.7   \\
4 &
  1.7   &
  10.5    \\
5 &
  0.005  &
  1.0   \\
6 &
  0.04   &
  3.4   \\
7 &
  0.15  &
  4.6   \\
8 &
  0.78  &
  6.3   \\
9 &
  0.77   &
  6.1   \\
10 &
   0.74  &
   5.6  \\
11 &
   --- &
   --- \\
12 &
   ---  &
   ---  \\
13 &
   ---  &
   --- \\
14 &
   0.31 &
   0.81  \\
15 &
   0.23  &
   0.56 \\
16 &
   0.28 &
   0.80  \\
17 &
   0.14 &
   0.64  \\
18 &
   0.60 &
   1.4 \\

\tableline
\end{tabular}
\end{center}
\tablecomments{Fluxes exclude the fitted eclipse component given in
Table 4.}
\end{table}

\clearpage
\noindent
{\large\bf Figure Captions}

\figcaption[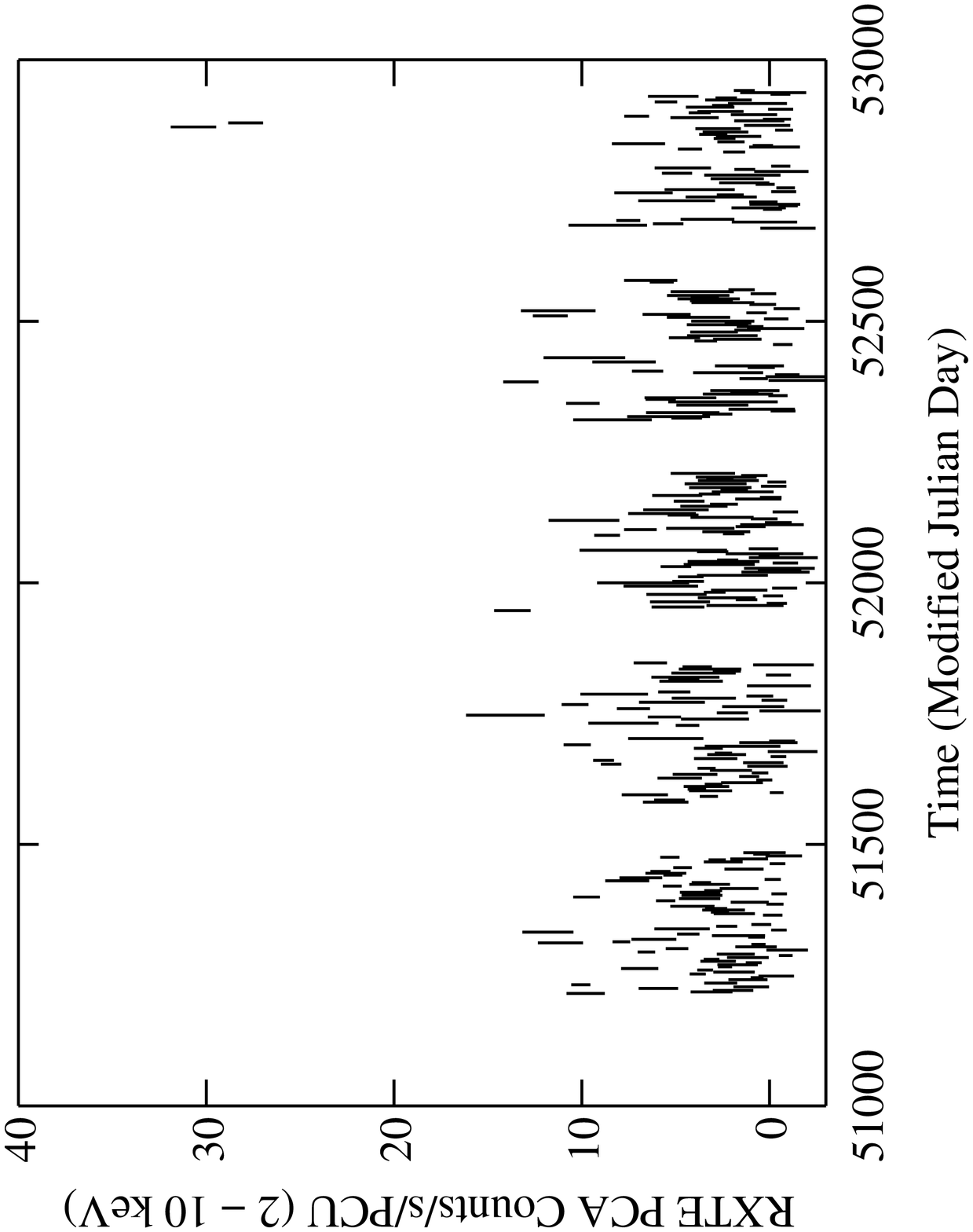]{The background subtracted light curve of \src\ obtained
with the PCA from scans over the Galactic center region.}

\figcaption[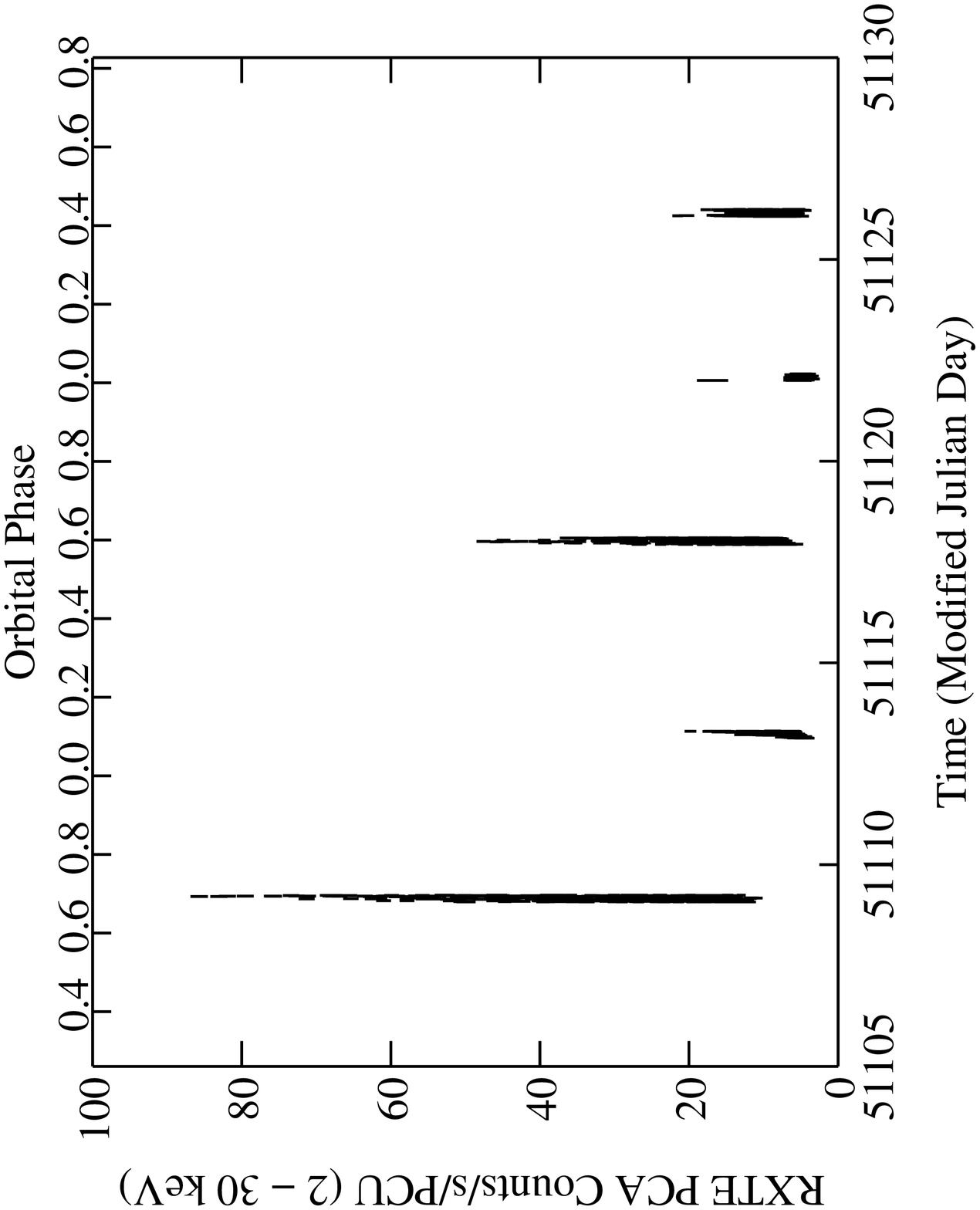]{The background subtracted light curve of \src\
obtained from pointed observations made with the PCA.}

\figcaption[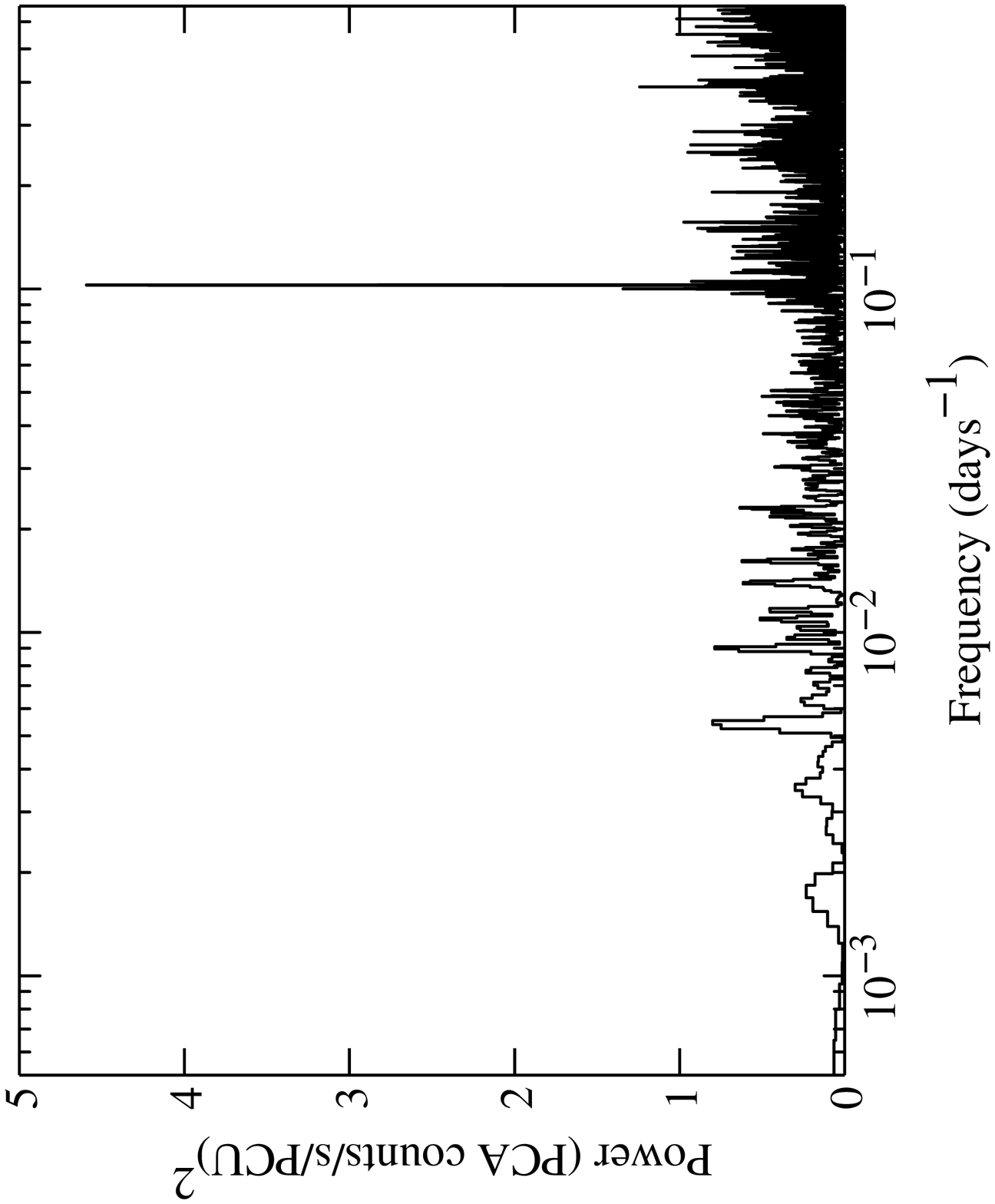]{The power spectrum of the PCA scan light curve
of \src.}

\figcaption[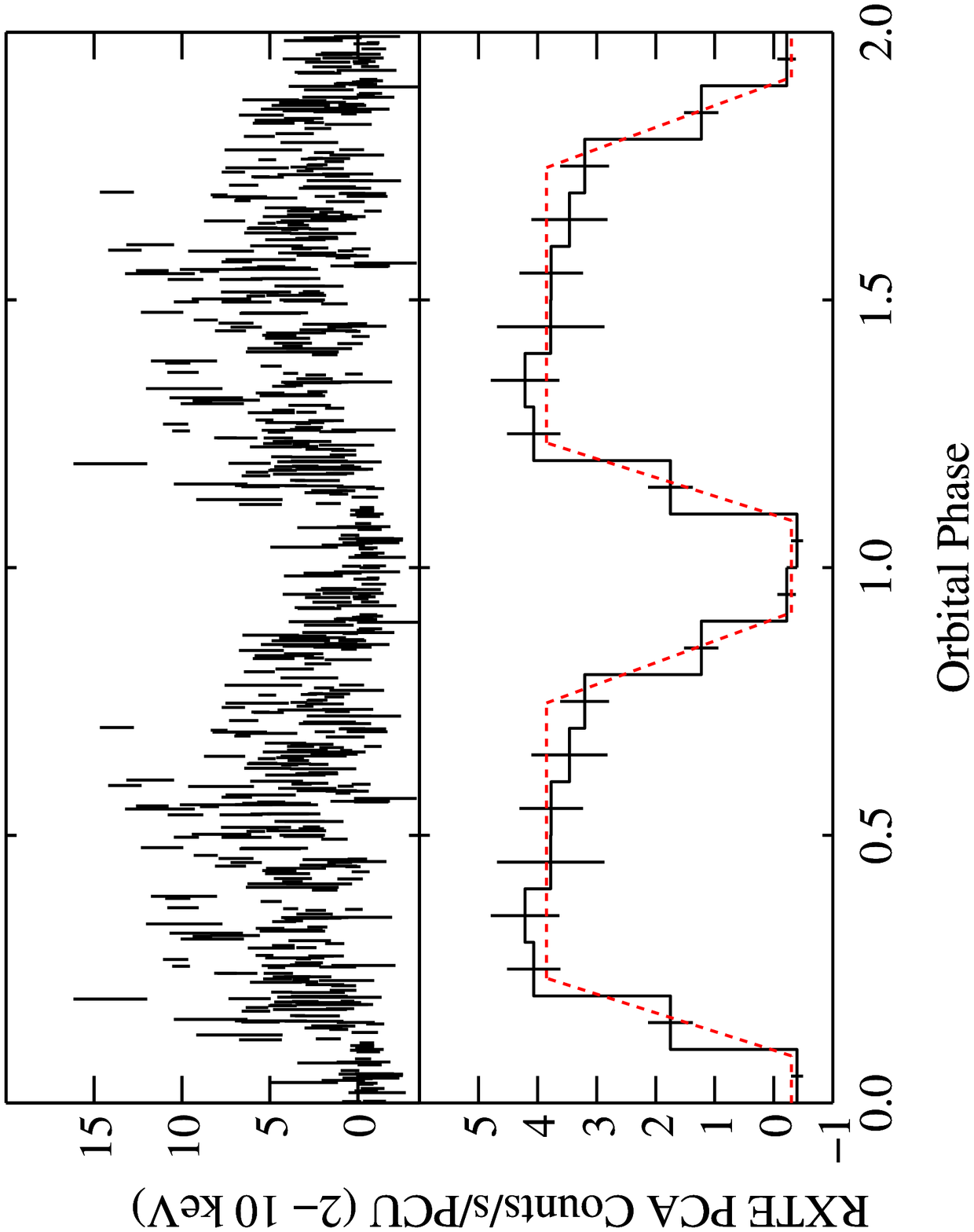]{The PCA scan light curve of \src\ folded
on the orbital period. The upper panel shows folded individual observations.
The lower panel shows a binned version of the upper panel together
with the eclipse fit given in Section 3.1.1 and Table 1 plotted as
a dashed line.}

\figcaption[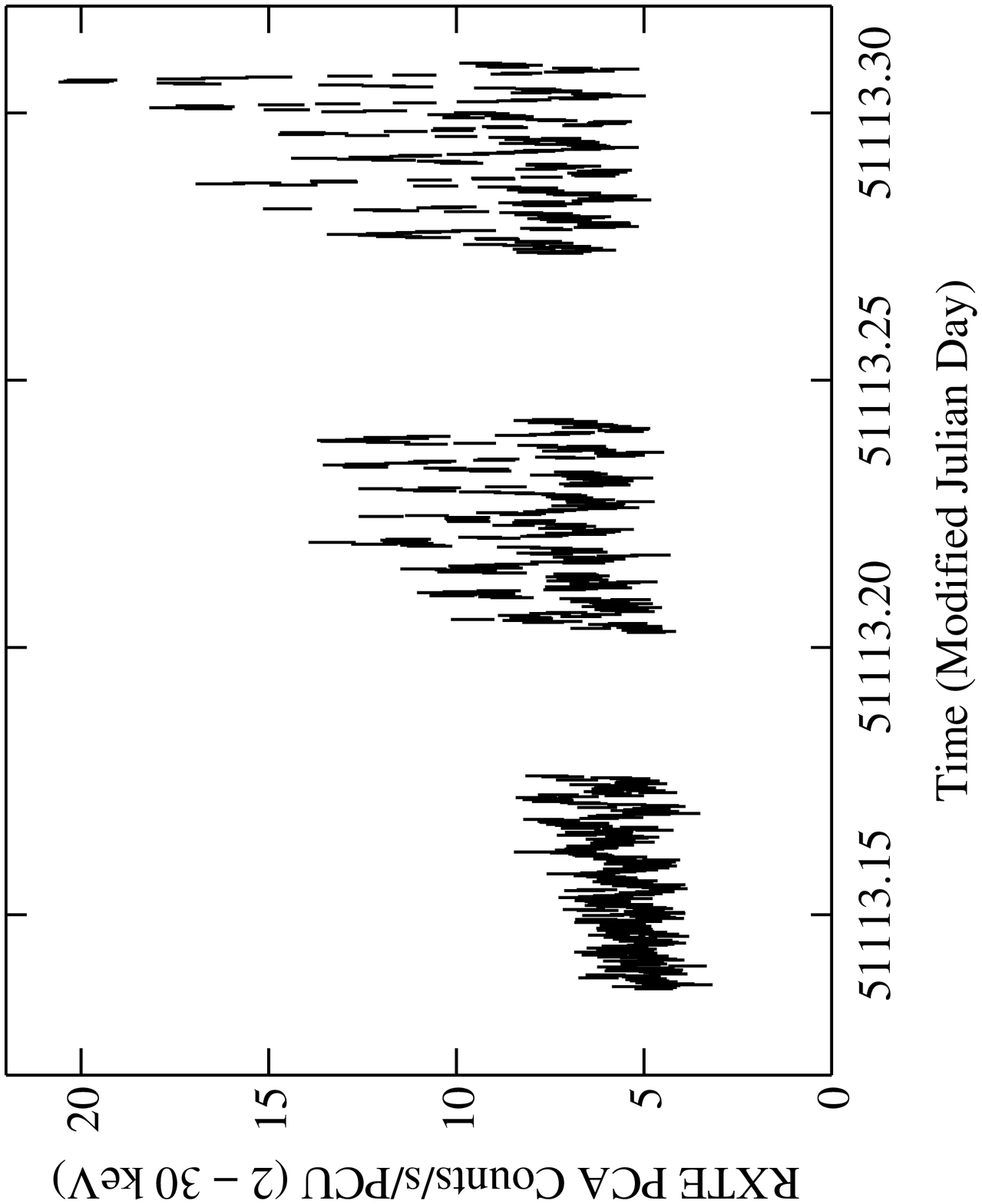]{Detail of the background subtracted light curve of \src\
obtained from pointed observations made with the PCA showing
an apparent egress from eclipse. The three data stretches shown
correspond to spectra 5, 6, and 7 which exhibit large
column densities.}

\figcaption[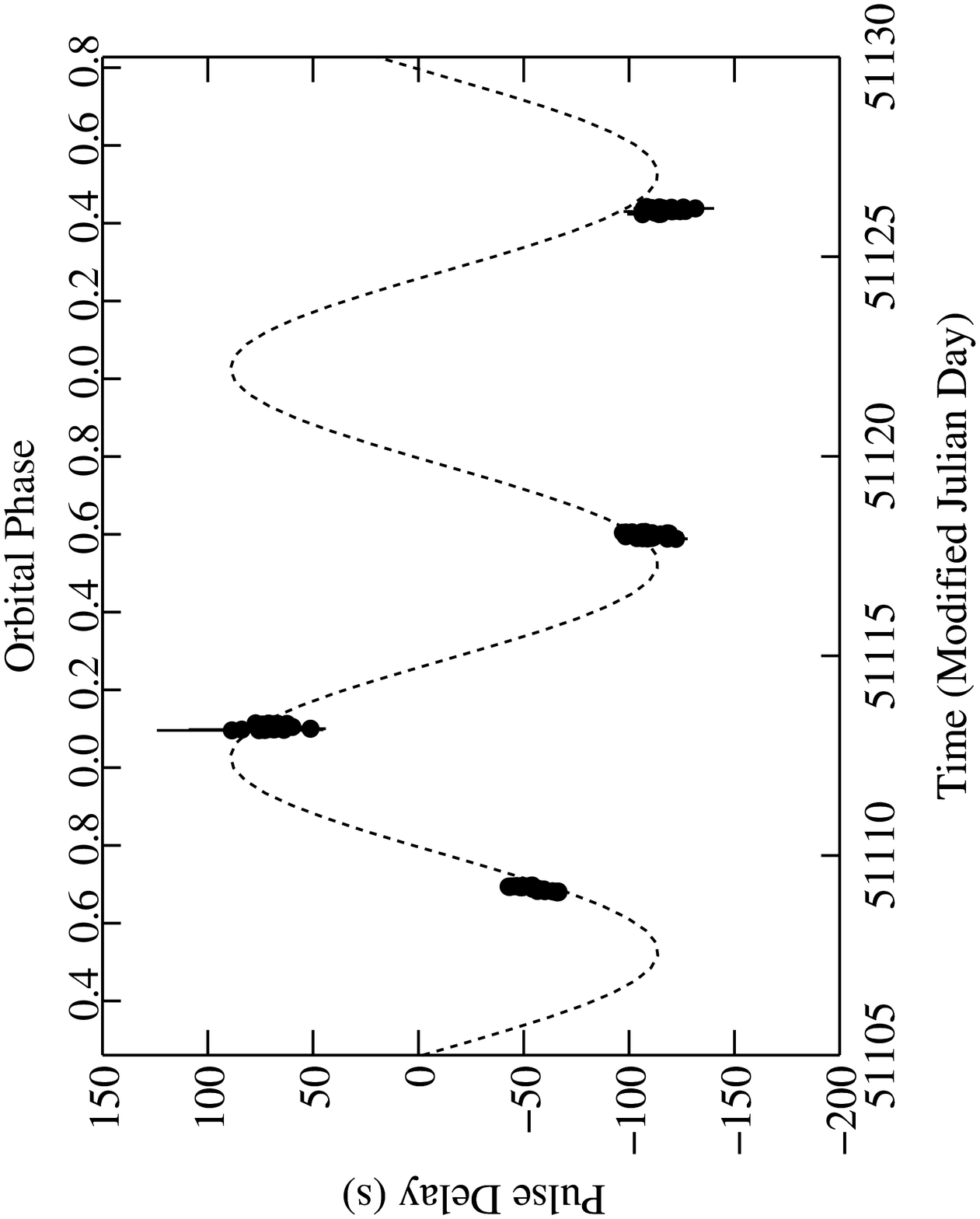]{The pulse delay curve for \src. The dashed line indicates
an example of a fitted circular orbit. Note that there is ambiguity between
several parameters in the pulse delay curve due to the poor sampling of the orbit.
Alternative delay curves are possible where non-contiguous sections of the delay curve
may be shifted by integral numbers of pulse periods.}

\figcaption[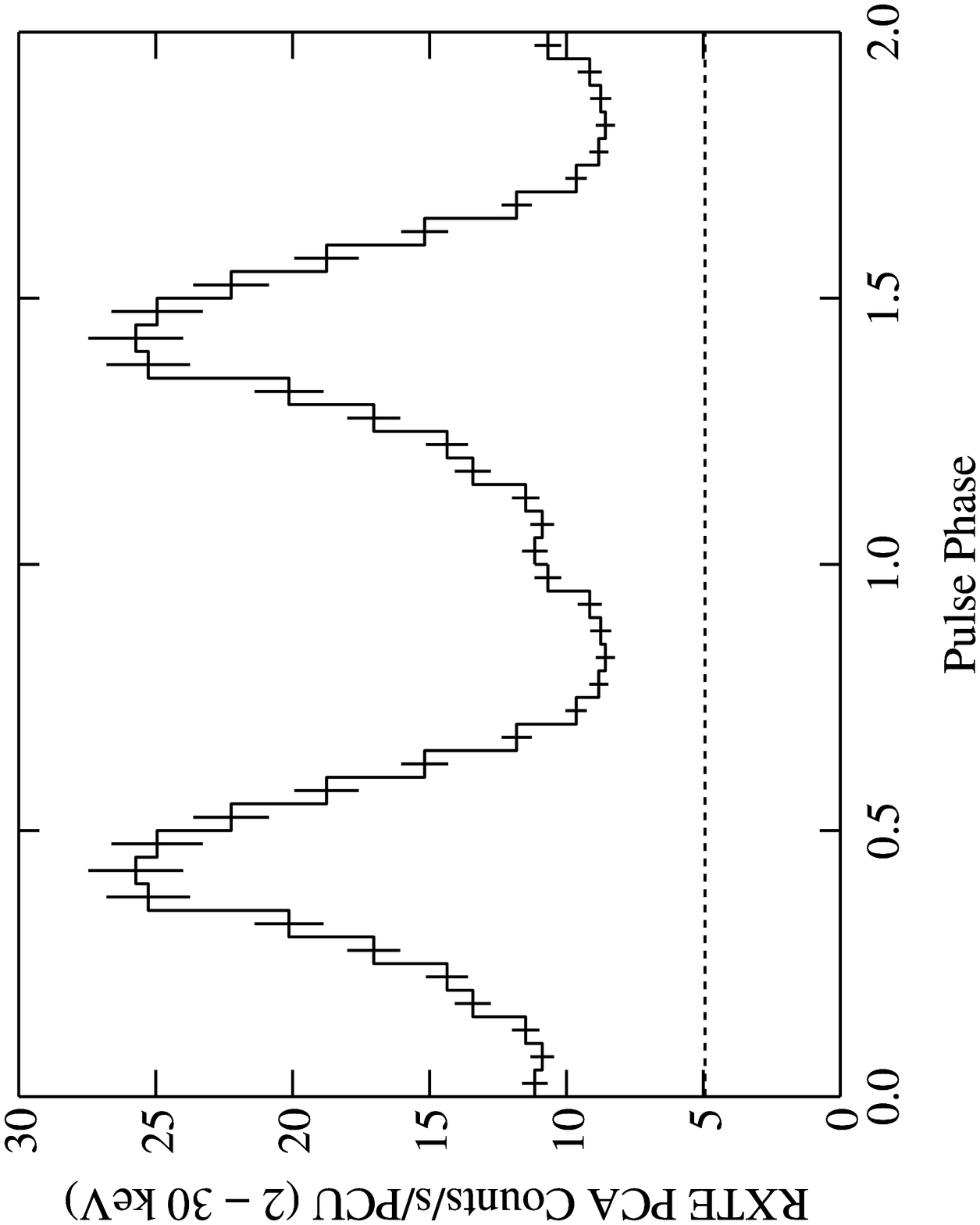]{The mean background subtracted pulse profile of \src\ obtained
with the PCA folded on a period of 413.308s. The dashed line indicates the count rate observed during
the apparent eclipse of the source.}


\figcaption[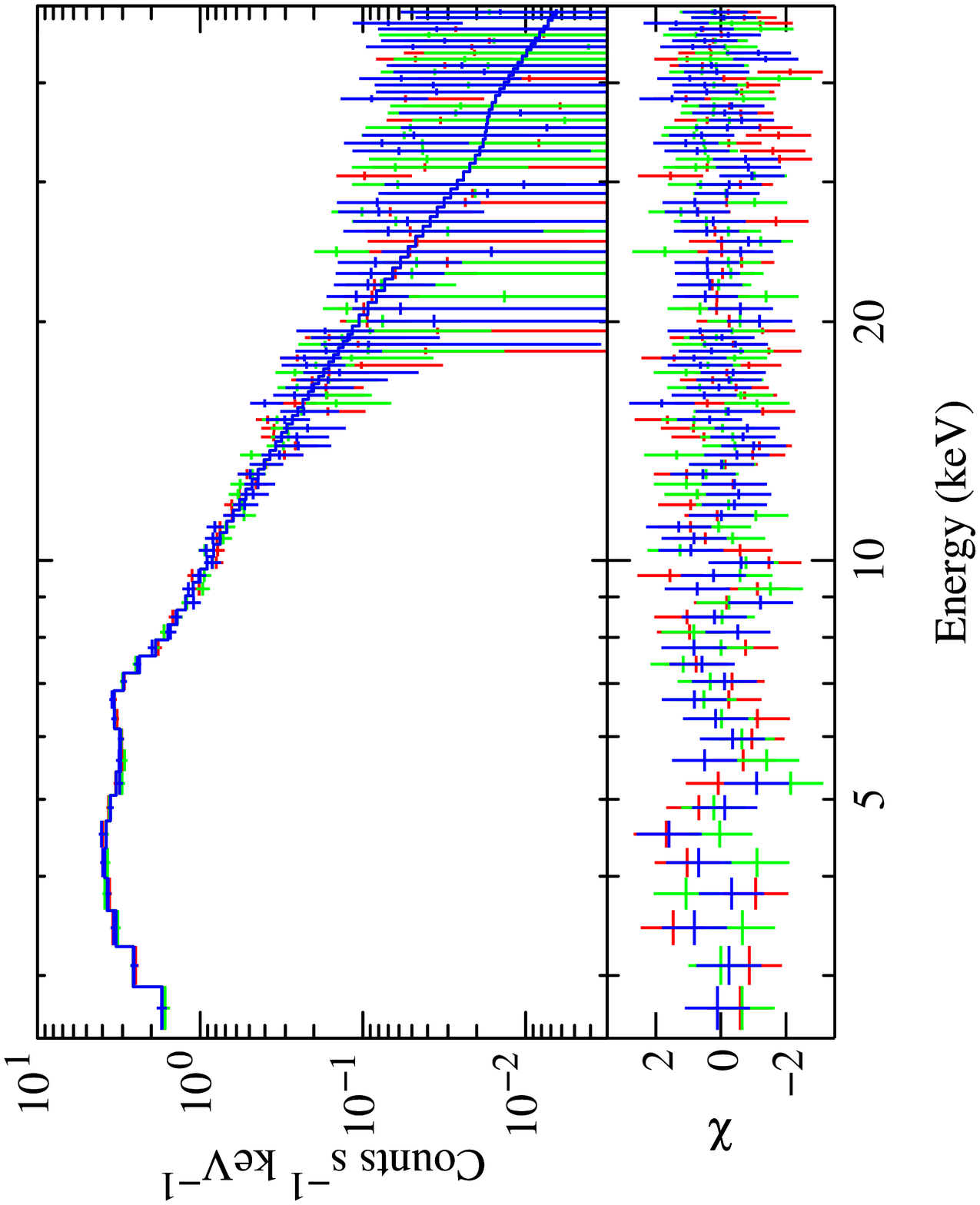]{Fits to the spectrum of \src\ during the eclipse.
Parameters are given in Table 4. Red lines show spectrum 11, green lines
show spectrum 12, and blue lines show spectrum 13.}

\figcaption[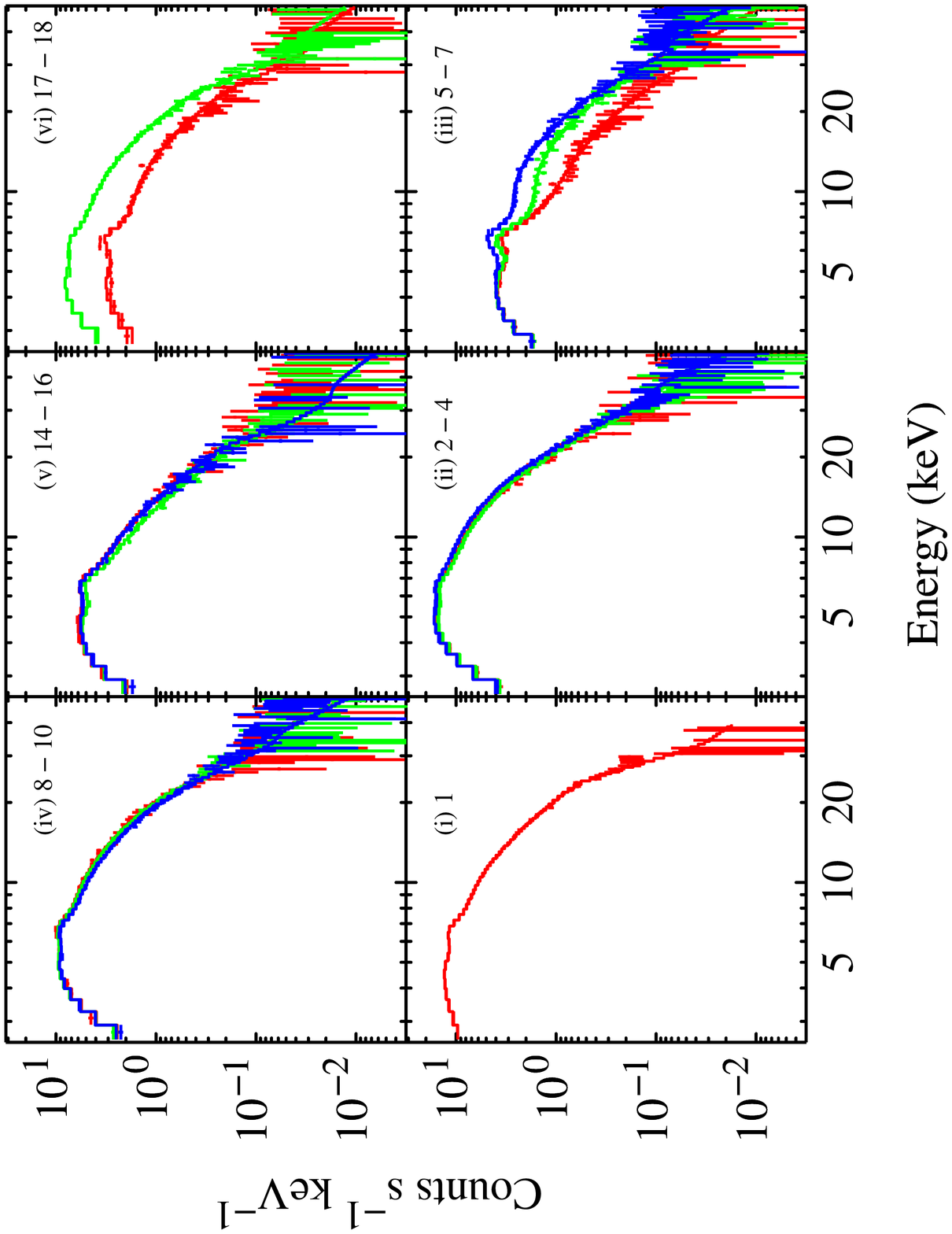]{Fits to the non-eclipse spectra of \src.
Parameters are given in Table 5. Each panel shows spectra that
were fit simultaneously. Panel (i): spectrum 1. Panel (ii):
spectra 2 (red), 3 (green), \& 4 (blue). Panel (iii): spectra 5 (red),
6 (green), \& 7 (blue). Panel (iv): spectra 8 (red), 9 (green)
and 10 (blue). Panel (v): spectra 14 (red), 15 (green), \& 16 (blue).
Panel (vi): spectra 17 (red) \& 18 (green).}

\figcaption[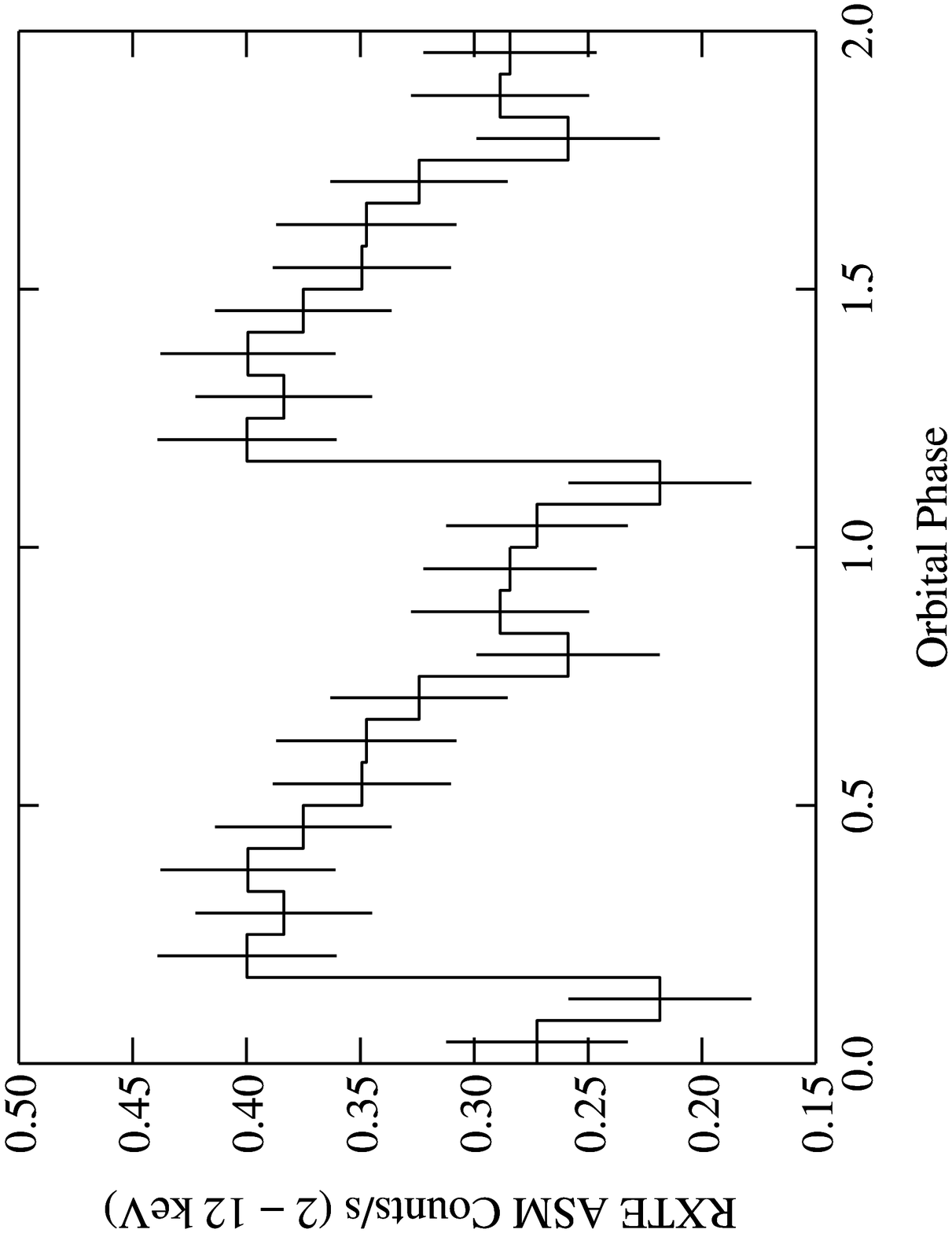]{The ASM light curve of \src\ folded
on the orbital period.}

\begin{figure}
\plotone{f1.eps}
\end{figure}

\begin{figure}
\plotone{f2.eps}
\end{figure}

\begin{figure}
\plotone{f3.eps}
\end{figure}

\begin{figure}
\plotone{f4.eps}
\end{figure}

\begin{figure}
\plotone{f5.eps}
\end{figure}

\begin{figure}
\plotone{f6.eps}
\end{figure}

\begin{figure}
\plotone{f7.eps}
\end{figure}

\begin{figure}
\plotone{f8.eps}
\end{figure}

\begin{figure}
\plotone{f9.eps}
\end{figure}

\begin{figure}
\plotone{f10.eps}
\end{figure}

\end{document}